\documentclass[useAMS,usenatbib]{mn2e}

\usepackage{amssymb}
\usepackage{amsfonts}
\usepackage{natbib}
\usepackage{epsfig}

\def\gs{\mathrel{\raise0.35ex\hbox{$\scriptstyle >$}\kern-0.6em 
\lower0.40ex\hbox{$\scriptstyle \sim$}}}
\def\ls{\mathrel{\raise0.35ex\hbox{$\scriptstyle <$}\kern-0.6em 
\lower0.40ex\hbox{$\scriptstyle \sim$}}}
\newcommand{\msun}{M_\odot}

\newcommand{\be}{\begin{equation}}
\newcommand{\ee}{\end{equation}}
\newcommand{\bea}{\begin{eqnarray}}
\newcommand{\eea}{\end{eqnarray}}

\begin{document}
\title[MOND plus classical neutrinos not enough for cluster lensing]{MOND 
plus classical neutrinos are not enough for cluster lensing}

\author[Natarajan \& Zhao] {Priyamvada 
Natarajan$^{1,2}$ and Hongsheng Zhao$^{3,4}$\\
$^1$ Department of Astronomy, Yale University, P. O. Box 208101, 
New Haven, CT 06511-208101, USA \\
$^2$ Department of Physics, Yale University, P. O. Box 208120, 
New Haven, CT 06520-208120, USA\\
$^3$ School of Physics and Astronomy, University of St. Andrews, 
North Haugh, St Andrews KY16 9SS Scotland\\
$^4$ Dark Cosmology Center, Copenhagen}

\maketitle

\begin{abstract}
Clusters of galaxies offer a robust test bed for probing the nature of
dark matter that is insensitive to the assumption of the gravity
theories. Both Modified Newtonian Dynamics (MOND) and General
Relativity (GR) would require similar amounts of non-baryonic matter
in clusters as MOND boosts the gravity only mildly on cluster
scales. Gravitational lensing allows us to estimate the enclosed mass
in clusters on small ($\sim\,20\,-\,50\,{\rm kpc}$) and large ($\sim$
several 100 kpc) scales independent of the assumptions of
equilibrium. Here we show for the first time that a combination of
strong and weak gravitational lensing effects can set interesting
limits on the phase space density of dark matter in the centers of
clusters. The phase space densities derived from lensing observations
are inconsistent with neutrino masses ranging from 2 - 7 eV, and
hence do not support the 2 eV-range particles required by MOND. To
survive, the plausible modifications for MOND may be either an additional 
degree of dynamical freedom in a co-variant incarnation or mass-varying
theories of neutrinos.
\end{abstract}


\section{Introduction}

The Newtonian Poisson equation, if sourced by purely ordinary baryonic
 matter, seriously under-predicts the accelerations seen in a wide
 range of scales. An extra source term, customarily known as either
 dark matter or a scalar field source in MOND, is needed for
 consistency (Zhao \& Famaey 2006; Famaey, Gentile, Bruneton \& Zhao
 2007). The precise nature of dark matter remains unknown within the
 Newtonian framework as well. The most viable dark matter candidate is
 a fermionic, neutral particle that condensed from the thermal bath of
 the early Universe (Kolb \& Turner 1990). While the detection of
 neutrinos confirmed the concept of a possible, particulate dark
 matter candidate, it did very little in closing the wide gap between
 observed matter density versus that what is required to explain
 lensing results and motions on galaxy scales. There remains the need
 for more exotic species of dark matter. In principle, galaxy scale
 observations ought to be sensitive to the free-streaming scale of
 these particles.

Another approach that has been followed is to constrain the phase
space densities of dark matter particles.  Studying the phase space
density of galaxy scale halos derived from observational data,
Sellwood (2000) noted that while the peak phase space density of dark
matter in galaxies is far from having a universal value, there does
appear to be a favoured scale of a few keV. The same keV mass scale
reappears in the recent discussions of the cores and dynamical
friction in dwarf spheroidals such as Ursa Minor (Kleyna et al. 2005)
and Fornax (Goerdt et al. 2006). The dark matter dominated Fornax
dwarf spheroidal has globular clusters orbiting at roughly $\sim$ 1
kpc from its centre.  Goerdt et al. (2006) argue that if the dark
matter halo hosting Fornax has a cuspy density profile, the globular
clusters would sink to the centre from their current positions within
a few Gyr, presenting a puzzle as to why they do indeed survive at the
present epoch. They show that this timing problem is alleviated by
adopting a cored dark matter halo. In that instance, using numerical
simulations and analytic calculations they argue that the sinking time
is many Hubble times; and the globulars would effectively halt at the
core radius of the dark matter halo. Using the current positions of
the globulars Goerdt et al. (2006) therefore conclude that the Fornax
dwarf spheroidal has a shallow inner density profile with a finite
core radius.  This immediately implies that the dark matter component
is warm, with an upper limit to its mass of $\sim\,0.5\,{\rm
keV}$. Such a warm dark matter candidate would suppress structure
formation on small scales alleviating another problem the so-called
substructure problem that seems to be endemic to Cold Dark Matter
(CDM) models.  On the other hand, the flux anomalies of
gravitationally lensed quasars argues for the existence of kpc clumps,
too dense for keV warm dark matter particles (Metcalf \& Zhao 2002;
Miranda \& Maccio 2007).
 
However, in competing gravity theories like MOND, the interpretation
of dwarf galaxy scale rotation curves is very different due to the
much larger MONDian gravity than predicted by Newtonian theory. With a
simple boost of gravity below a scale $a_0\,\sim\,10^{-8}$m/s$^2$, the
need for dark matter on dwarf galaxy scales becomes much weaker if we
assume MOND (Famaey, Gentile, Bruneton \& Zhao 2007 and references
therein, Sanders \& McGaugh 2002). Part of the reason that MOND is
able to mimic CDM effectively is that there is a common acceleration
scale $g\,\sim\,a_0$ in the dark matter cusp of the Navarro, Frenk,
White profile (Xu, Wu \& Zhao 2007), which appears on galaxy to galaxy
cluster scales.  This is exactly the scale on which MOND can
supplement ordinary gravity, so it is not surprising that MOND and CDM
often give comparable fits to data. However, there are still a few
tough challenges for MOND (Klypin \& Prada 2007; Famaey, Bruneton \&
Zhao 2007) even on these scales.  Additionally, Zhao (2005) noted that
MOND would require globular clusters and dwarf galaxies to have the
same size tidal radius, which appears to contradict current
observations.  So while there are ambiguities on galaxy scales,
cluster scales are more promising to discriminate between the two
theories as the effects of MOND are expected to be mild.  This is due
to the fact that MOND boosts the gravitational constant up only by a
mild factor $1/\mu(x)$, where $\mu(x) = {x \over 1+x} \sim 0.5$ in
clusters of galaxies (Wu, Zhao, Famaey, Gentile, Tiret, Combes, Angus
\& Robin 2007). The boosting is much larger in dwarf galaxies, by a
factor of $\sim 11$, where $x \sim 0.1$. In short, the evidence for DM
on galaxy scales is weak in the context of alternatives like
MOND. However, as we show below, more robust tests definitely derive
from lensing in clusters of galaxies, in particular, the potent
combination of strong and weak lensing observations.
 
Clusters of galaxies are the most massive and recently assembled
structures in the Universe. In the context of the hierarchical growth
of structure in a cold dark matter dominated Universe, clusters are
the repository of copious amounts of the dark matter. Gravitational
lensing, predicted by Einstein's theory of General Relativity, is the
deflection of light rays from distant sources by foreground mass
structures is now detected in over a 100 clusters. Dramatic strong
lensing occurs when there is a rare alignment of background sources
with the dense central region of a foreground cluster. This produces
highly distorted, magnified and multiple images of a single background
source (Schneider, Ehlers \& Falco 1992). However, more commonly, the
observed shapes of background sources viewed via a foreground cluster
lens are systematically elongated, in the so-called weak lensing
regime. Coupling strong and weak lensing offers the most reliable
probe of the distribution of dark matter on various cosmic scales
(Blandford \& Narayan 1992; Mellier 2002; Schneider, Ehlers \& Falco
1992).  In particular, the combination of data from these two regimes
offers an unprecedented insight into the detailed mass distribution of
clusters (Natarajan, Kneib, Smail \& Ellis 1998; Natarajan, Kneib \&
Smail 2002; Bradac et al. 2006; Jee et al. 2007). The lensing
distortion in the shapes of background galaxies viewed through
fore-ground mass distributions is independent of the dynamical state
of the lens, therefore, unlike other methods for mass estimation there
are fewer biases in lensing mass determinations.

Strong lensing studies of the inner regions of several clusters
indicate that the dark matter distribution can be represented by the
combination of a smoothly distributed, extended component and
smaller-scale clumps or subhaloes associated with luminous galaxies
(Kneib et al. 1996; Natarajan \& Kneib 1997; Natarajan, Kneib, Smail
\& Ellis 1998). The smooth component has been detected using weak
lensing techniques out to the turn-around radius (typically of the
order of several Mpc) in clusters (Kneib et al. 2003; Broadhurst et
al. 2005). The lensing derived density profile of the smooth
component, and its agreement with profiles computed from high
resolution numerical simulations of structure formation in the
Universe is currently well studied (Navarro, Frenk \& White 1997;
Navarro et al. 2004; Sand et al. 2004).  In addition, the granularity
of the dark matter distribution associated with individual galactic
subhaloes holds important clues to the growth and assembly of clusters
(Tyson, Kochanski \& dell'Antonio 1998; Broadhurst, Huang, Frye \&
Ellis 2000; Limousin et al. 2007; Jee et al. 2007).
 
We exploit the technique of galaxy-galaxy lensing, which was
originally proposed as a method to constrain the masses and spatial
extents of field galaxies (Brainerd, Blandford \& Smail 1996). The
methodology has since been extended and developed to apply inside
clusters (Natarajan \& Kneib 1996; Natarajan et al. 1998; 2002a;
Natarajan, De Lucia \& Springel 2007). Constraints on the masses of
subhaloes associated with galaxies in clusters is also now available
for several clusters.
 
The outline of this paper is as follows: in Section 2 we present the
constraints obtained to date on MOND from clusters, in Section 3 we
describe our method to derive the central densities of clusters and
cluster galaxies, implications thereof for MOND are in Section 4 and
we conclude with a discussion in Section 5.
 
\section{Warming up to MOND predictions in clusters}

Modified Newtonian Dynamics (MOND), was proposed by Milgrom (1983) as
an alternative to Newtonian gravity, to explain galactic dynamics
without the need for dark matter. Although current cosmological
observations point to the existence of vast amounts of non-baryonic
dark matter in the Universe, it is interesting and important to
explore other alternatives. While this component is found to be
distributed on a range of length scales, clusters of galaxies seem to
be sites that are in fact dominated by dark matter at almost all
radii.

In MOND the gravitational force at large distances and small
accelerations is modified when the acceleration is lower than a
critical value, defined to be $g_0\,=\,1\,\times\,10^{-8}$cm/s$^2$.
With one free parameter, namely the mass-to-light ratio, this
formulation can explain rather well the rotation curves notably of low
surface brightness galaxies (McGaugh 2004; 2005; Gentile et al. 2007;
McGaugh et al. 2007). Further tests of the MOND theory are needed in
order to understand if there is a fundamental need for such a
modulation of gravitational forces.

Recently Bekenstein (2004) proposed a relativistic formulation of
MOND, called TeVeS. This enabled the calculation of relativistic
phenomena and in particular, the deflection of light rays propagating
in a MOND Universe. With such a formulation in hand, the theory can be
now be tested using the plethora of gravitational lensing observations
currently available. Zhao et al. (2006) have examined the implications
for galaxy scale lenses and find that while many candidates in the
CASTLES survey are compatible with MOND lensing, there are
outliers. Clusters of galaxies offer a more powerful probe as a range
of lensing phenomena occur in them.

The Bullet cluster (1E 0657-56) opened an interesting debate about the
nature of dark matter in clusters. The clear separation of the lensing
shear signal from the X-ray gas signal implies some form of dark
matter in general relativity (Clowe et al. 2006; Bradac et al. 2006)
as well as in MOND. Angus et al. (2007) showed that the data can be
reconciled with $2\,{\rm eV}$ neutrinos, a neutrino mass limit that is
allowed by current beta-decay experiments. Thus, they argue that this
alleviates the need for more exotic dark matter in MOND and does not
offer concrete proof for the existence of cold dark matter as the dark
matter density in these clusters is very low, and can be easily
explained by the phase space density of neutrinos. Nevertheless,
several preprints since (Angus, Famaey \& Boute 2007) have started to
reveal the inadequacy of neutrinos as a plausible constituent for the
Bullet Cluster within the MOND framework. And as we show here the
central densities for clusters and cluster galaxies estimated by
combining strong and weak lensing constraints out to several hundred
kpc cannot be explained by ${2\,{\rm eV}}$ neutrinos.

In a recent preprint, Takahashi \& Chiba (2007) have explored the
implications of weak lensing data of clusters for MOND. Using
published weak lensing data for 3 Abell clusters and 42 SDSS clusters,
they conclude that MOND cannot explain the data unless a dark matter
halo is added. They find that dark matter is required as it cannot be
accounted for with neutrinos with masses less than 2 eV 
\footnote{Note that Cl\,0024+16 one of the clusters studied here was
part of the Takahashi \& Chiba sample. However, the only constraint
they employ is from the weak lensing data that does not provide a
calibrated mass distribution.}. In earlier work, a massive neutrino
with a mass of $\sim$ 2 eV was invoked as dark matter to explain
observational data (Sanders 2003; Skordis et al. 2006). In our work,
reported here we consider the detailed mass distribution in 6 clusters
spanning a redshift range of $z = 0.2 - 0.6$ with very well calibrated
lensing models. Weak lensing observations alone do not give an
absolute mass calibration, another mass estimator is needed for the
normalization. However, combining strong and weak lensing data using
measured spectroscopic redshifts for the multiple images enables us to
construct calibrated mass models.

Combining strong and weak lensing data from the Hubble Space
Telescope's (HST) Wide-Field Planetary Camera (WFPC-2), with a large
complement of ground based spectroscopy, we have constructed high
resolution mass models for Abell 2218, Abell 2390, AC\,114,
Cl\,2244-02, Cl\,0024+16 and Cl\,0054-27. This enables us to compute
the central density of these clusters and that of the typical subhalo
that hosts an early-type $L^*$ galaxy in these clusters.

Note that while in the following sections we will only construct
models with Newtonian-Einsteinian gravity, our results are applicable
to MOND due to the simple fact that all particles accelerate inside
clusters with $g\,\sim\,(0.5a_0\,-\,3a_0)$ in general.\footnote{At the
Einstein ring radius of $r_E\,\sim\,50$ kpc, and $\sigma\,\sim\,500$ km/s,
the acceleration can be estimated by: $g\,\sim\,2\,\sigma^2/r_E\,\sim\,3
a_0$. So in the region of interest we are in moderate or strong
gravity, hence MOND effects are mild.} Therefore, MOND effects are
always mild, within a factor of two at most. Wu et al. (2007) showed
that the MOND gravity around a galaxy inside a cluster should have
nearly Newtonian or Keplerian behavior. Hence we will model lensing in
the Newtonian-Einsteinian framework, but extrapolate our conclusions
to MOND.

\section{Construction of mass models from lensing data}

In this section, we briefly outline the method used to derive
constraints on the mass distribution of clusters and galaxies in
clusters. The mass distribution in clusters is partitioned into a
large scale smooth component of dark matter and small scale subhaloes
that are associated with the locations of bright cluster galaxies.

To quantify the lensing distortion induced, the large scale smooth
component and the individual galaxy-scale halos are modeled
self-similarly using the Pseudo Isothermal Elliptical Mass
Distribution (PIEMD: Kassiola \& Kovner 1993) profile with,
\begin{eqnarray}
\Sigma(R)\,=\,{\Sigma_0 r_0  \over {1 - r_0/r_t}}
({1 \over \sqrt{r_0^2+R^2}}\,-\,{1 \over \sqrt{r_t^2+R^2}}),
\end{eqnarray}
with a model core-radius $r_0$ and an outer cut radius $r_t\,\gg\,
r_0$. The coordinate $R$ is a function of $x$, $y$ and the ellipticity, 
\bea
R^2\,=\,({x^2 \over (1+\epsilon)^2}\,+\,{y^2 \over (1-\epsilon)^2})\,;
\ \ \epsilon= {a-b \over a+b}, 
\eea 
The mass enclosed within an aperture radius $R$ for the $\epsilon = 0$ 
model is given by: 
\be 
M(R)={2\pi\Sigma_0
r_0 \over {1-{{r_0} \over {r_t}}}}
[\,\sqrt{r_0^2+R^2}\,-\,\sqrt{r_t^2+R^2}\,+\,(r_t-r_0)\,].  
\ee 
The total mass $M$, is finite $M\,
\propto \,{\Sigma_0} {r_0} {r_t}$. 
The shear is:
\bea
\gamma(R)\,&=&\,\nonumber \kappa_0[\,-{1 \over \sqrt{R^2 + r_0^2}}\, +\,{2 \over
R^2}(\sqrt{R^2 + r_0^2}-r_0)\,\\ \nonumber &+&\,{1 \over {\sqrt{R^2 +
r_t^2}}}\,-\, {2 \over R^2}(\sqrt{R^2 + r_t^2} - r_t)\,].\\ 
\eea
In order to relate the light distribution in cluster galaxies to key
parameters of the mass model of subhaloes, we adopt a set of physically
motivated scaling laws derived from observations (Brainerd et al.\
1996; Natarajan \& Kneib 1997; Limousin et al. 2004): 
\begin{eqnarray}
{\sigma_0}\,=\,{\sigma_{0*}}({L \over L^*})^{1 \over 4};\,\,
{r_0}\,=\,{r_{0*}}{({L \over L^*}) ^{1 \over 2}};\,\,
{r_t}\,=\,{r_{t*}}{({L \over L^*})^{\alpha}}.
\end{eqnarray}
The total mass $M$ enclosed within an aperture $r_{t*}$ and
the total mass-to-light ratio $M/L$ 
then scale with the luminosity as follows for the early-type galaxies:
\begin{eqnarray}
M_{\rm ap}\,\propto\,{\sigma_{0*}^2}{r_{t*}}\,({L \over L^*})^{{1 \over
2}+\alpha},\,\,{M/L}\,\propto\,
{\sigma_{0*}^2}\,{r_{t*}}\left( {L \over L^*} \right )^{\alpha-1/2},
\end{eqnarray}
where $\alpha$ tunes the size of the galaxy halo. These scaling laws
are empirically motivated by the Faber-Jackson relation for early-type
galaxies (Brainerd, Blandford \& Smail 1996). For late-type cluster
members when the data is available (at the present time only for the
cluster Cl\,0024+16), we use the analogous Tully-Fisher relation to
obtain scalings of $\sigma_{0*}$ and ${r_{t*}}$ with luminosity. The
empirical Tully-Fisher relation has significantly higher scatter than
the Faber-Jackson relation (see Courteau et al. 2007 \&
 Jorgensen et al. 2006). In this analysis we do not take the
scatter into account while employing these scaling relations. We
assume these scaling relations and recognize that this could
ultimately be a limitation but the evidence at hand supports the fact
that mass traces light efficiently both on cluster scales (Kneib et
al. 2003) and on galaxy scales (McKay et al. 2001; Wilson et
al. 2001). The details of the redshift distribution and intrinsic
ellipticity distribution assumed for this analysis (and for most
lensing analysis in fact) are described in detail in Natarajan et
al. (2007). While the core radius of the large scale smooth components
is constrained from observations, the core radii of the individual
cluster galaxies cannot be constrained with current data. Therefore,
in the modeling, we fix the core radius of a dark matter subhalo that
hosts an $L^*$ galaxy to be $0.1\,{\rm kpc}$.  This assumption will
not be of consequence in the determination of the central density in
cluster galaxies as discussed below.

Parameters that characterize both the global components and the
perturbers are optimized, using the observed strong lensing features -
positions, magnitudes, geometry of multiple images and measured
spectroscopic redshifts, along with the smoothed shear field as
constraints. With the parameterization presented above, we optimize
and extract values for the central velocity dispersion and the
aperture scale $(\sigma_{0*}, r_{t*})$ for a subhalo hosting a
fiducial $L^*$ cluster galaxy. We note here that as argued above
MOND is unimportant inside these extremely dense
and massive lensing clusters therefore cannot be used as a criterion
to question these scaling relations. The scaling relation used in this
paper the Faber-Jackson relation is well established observationally in
clusters and in lensing clusters. The Faber-Jackson relation is a projection
of the Fundamental Plane and this is established observationally in clusters
over a range of redshifts (some recent references Jorgensen et al. 2006;
Fritz et al. 2005; Holden et al. 2005; Pahre et al. 1998). The Tully-Fisher 
relation is also an empirical relation detected out to these redshifts 
(see Courteau et al.  2007 for the most recent data).

Maximum-likelihood analysis is used to obtain significance bounds on
these fiducial parameters that characterize a typical $L^*$ subhalo in
the cluster. The likelihood function of the estimated probability
distribution of the source ellipticities is maximized for a set of
model parameters, given a functional form of the intrinsic ellipticity
distribution measured for faint galaxies. For each `faint' galaxy $j$,
with measured shape $\tau_{\rm obs}$, the intrinsic shape $\tau_{S_j}$
is estimated in the weak regime by subtracting the lensing distortion
induced by the smooth cluster models and the galaxy subhaloes,
\begin{eqnarray}
\tau_{S_j} \,=\,\tau_{\rm obs_j}\,-{\Sigma_i^{N_c}}\,
{\gamma_{p_i}}\,-\, \Sigma_n\,\gamma_{c}, 
\end{eqnarray}
where $\Sigma_{i}^{N_{c}}\,{\gamma_{p_i}}$ is the sum of the shear
contribution at a given position $j$ from $N_{c}$ perturbers. This
entire inversion procedure is performed numerically using code
developed that builds on the ray-tracing routine {\sc lenstool}
written by Kneib (1993){\footnote{This software is publicly available
at http://www.oamp.fr/cosmology/lenstool/}}. This machinery accurately
takes into account the non-linearities arising in the strong lensing
regime. Using a well-determined `strong lensing' model for the
inner-regions along with the shear field and assuming a known
functional form for $p(\tau_{S})$ the probability distribution for the
intrinsic shape distribution of galaxies in the field, the likelihood
for a guessed model is given by,
\begin{eqnarray}
 {\cal L}({{\sigma_{0*}}},{r_{t*}}) = 
\Pi_j^{N_{gal}} p(\tau_{S_j}),
\end{eqnarray}
where the marginalisation is done over $(\sigma_{0*},r_{t*})$.  We
compute ${\cal L}$ assigning the median redshift corresponding to the
observed source magnitude for each arclet. The best fitting model
parameters are then obtained by maximizing the log-likelihood function
$l$ with respect to the parameters ${\sigma_{0*}}$ and ${r_{t*}}$.
Note that the parameters that characterize the smooth component are
also simultaneously optimized. The results of this analysis for our
sample of clusters, i.e. values of $(\sigma_{0*},r_{t*})$ are
presented in earlier papers (Natarajan, Kneib \& Smail 2002; Natarajan
et al. 2007).

In summary, the basic steps of our analysis therefore involve lens
inversion, modeling and optimization, which are done using the {\sc
lenstool} software utilities (Kneib 1993). These utilities are used to
perform the ray tracing from the image plane to the source plane with
a specified intervening lens. This is achieved by solving the lens
equation iteratively, taking into account the observed strong lensing
features, positions, geometry and magnitudes of the multiple images.
We also include a constraint on the location of the critical line
(between 2 mirror multiple images) to tighten the optimization. In
addition to the likelihood contours, the reduced $\chi^2$ for the
best-fit model is also found to be robust.

In addition, our reconstructions enable us to derive the mass
function of dark matter subhaloes inside these clusters. We find very good 
agreement between the mass functions predicted by the $\Lambda$CDM
model derived from high resolution cosmological simulations and those
computed via the above method from lensing observations. More details
on this comparison can be found in Natarajan \& Springel (2004);
Natarajan, De Lucia \& Springel (2007) and Natarajan et al. (2007).
So it is interesting to note that the lensing observations are in
consonance with the predictions of the concordance cosmological model
and require dark matter.

\begin{table*}
\begin{center}
\begin{tabular}{lcccccc}
\hline\hline\noalign{\smallskip} ${\rm Cluster}$& $z$ &
${\sigma_{0\ast}}$&${r_{t\ast}}$&  $M/L_V$  &
${\rho_{\rm clus}(r=r_0)}$ \\ & &(km\,s$^{-1}$)
& (kpc) & $(M_{\odot}/L_{\odot}$) & (10$^{-3}$ $\msun$ pc$^{-3}$)\\ 
\noalign{\smallskip}
\hline \noalign{\smallskip} {A\,2218} & ${0.17}$ & ${180\pm10}$ &
${40\pm12}$ & ${5.8\pm1.5}$ & {3.95} \\

{A\,2390} & ${0.23}$ & ${200\pm15}$ & ${18\pm5}$ &
${4.2\pm1.3}$ & {16.95} \\

{AC\,114}  & ${0.31}$ & ${192\pm35}$ & ${17\pm5}$ &
${6.2\pm1.4}$ & {9.12} \\ 

{Cl\,2244$-$02} & ${0.33}$ & ${110\pm7}$ & ${55\pm12}$ &
${3.2\pm1.2}$ & {3.52} \\

{Cl\,0024+16} & ${0.39}$ & ${125\pm7}$ & ${45\pm5}$ &
${2.5\pm1.2}$ & {3.63} \\

{Cl\,0054$-$27} & ${0.57}$& ${230\pm18}$ & ${20\pm7}$ &
${5.2\pm1.4}$ & {15.84} \\
\noalign{\smallskip}
\hline
\end{tabular}
\end{center}
\caption{Parameters that define the mass models of the subhaloes for
the lensing clusters. For each cluster the central velocity dispersion
of a subhalo that hosts an $L^*$ galaxy ($\sigma_{0*}$ in km/sec), the
aperture radius $r_{t*}$ (in kpc), the stellar mass-to-light ratio in
the V-band interior to $r_{t*}$  in solar units and the density of the large scale
cluster component $\rho_{\rm clus}(r = r_0)$ evaluated at the core
radius $r_0$.}
\end{table*}

For the large scale cluster, we derive the central density from total
mass within a few hundred kpc, well outside the Einstein radius. We
emphasize here that the lensing analysis also provides an estimate of
the total mass enclosed within $r_{t*}\,\sim\,20\,-\,50$ kpc for
subhaloes, well outside the typical Einstein radius of $\sim\,1\,-\,5$
kpc. It is within the Einstein radius that the contribution of baryons
dominates.  Since our estimates probe the mass well beyond the
Einstein radius, the bulk of the mass detected here is dark matter on
both cluster and cluster galaxy scales as we demonstrate below.

\section{Estimating central densities of clusters and galaxies in clusters 
from lensing}

The mass models are well calibrated and do not suffer from the
ambiguity of the mass-sheet degeneracy as more than 2 sets of multiple
images with measured redshifts are used in constraining the mass of
each cluster. The best-fit mass models enable us to compute the
central density for the clusters. For the PIEMD model:
\begin{eqnarray}
\rho(r)_{\rm clus} = \frac{\rho_0}{(1 + \frac{r^2}{r_0^2})\,(1 + \frac{r^2}{r_t^2})}
\end{eqnarray}
where $r_0$ is the core radius and $r_t$ is the outer truncation
radius as before in eqns. (1 - 4), derived from the best-fit mass
model. The core radius is an additional parameter in these models,
which is of consequence in the computation of central densities.  For
the large scale cluster component the core-radius $r_0$ is also
optimized in the likelihood analysis. The best-fit values for the core
radii for these clusters are listed in Table~1. The error bars in the
calculated central density (plotted in Figure~1) are derived by
propagating the errors on the quantities shown in Table~1 and
correspond to 3-$\sigma$ error bars. We note here that due to effects 
like the anisotropy $\sigma$ can be uncertain by a factor of 0.7 - 1.4
and the computed density by factor of 2. 

\begin{figure}
\begin{center}
\includegraphics[height=9cm,width=9cm]{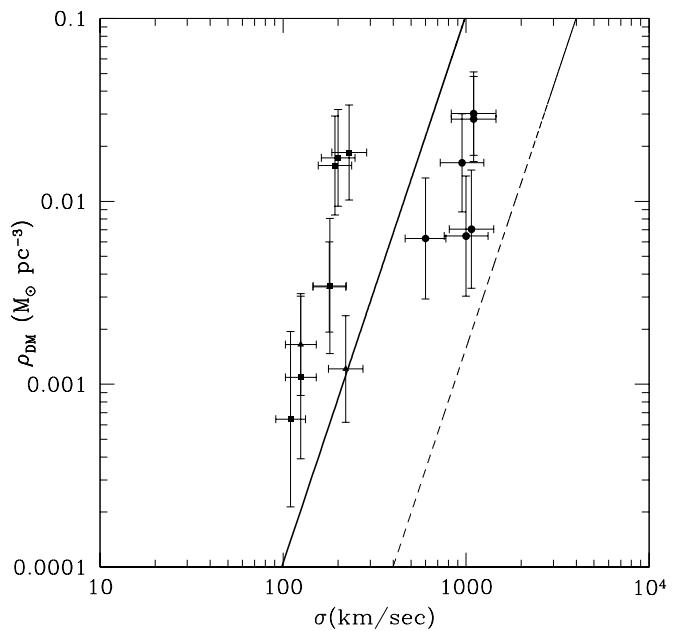}
\caption{The central densities of clusters and cluster galaxies
derived from lensing observations. Strong and weak gravitational
lensing are combined to obtain constraints on the dark matter
subhaloes associated with cluster galaxies. The cluster data points
are shown as solid circles, the solid squares and solid triangle are
values derived from the aperture masses derived from galaxy-galaxy
lensing studies, for early-type cluster galaxies and late-types
respectively. Note that the baryonic contribution has been subtracted
in the points plotted above. Lines show the maximum phase space
density for 2 species of sterile neutrinos with masses $7\,{\rm eV}$
and $6.9\,{\rm eV}$ + 3 species of ordinary neutrinos with masses of
$0.08\,{\rm eV}$ (solid line) and for only 3 species of ordinary
neutrinos (dashed line) with a mass of $2\,{\rm eV}$ each. The dashed
line is for $M_{\nu}^4 = 4.39$ and the solid line is for $M_{\nu}^4 =
291.73$. The central densities $\rho_0 = \rho(r = 0)$ are plotted for
the clusters and the averaged central density $<\rho>$ is plotted for
the cluster galaxies.}
\end{center}
\end{figure}

For the cluster galaxies, the likelihood method constrains the total
mass enclosed ($M_{\rm ap} \sim \sigma_{0*}^2\,r_{t*}$) within an
aperture $r_{t*}$.  We compute the average smoothed density simply
using:
\begin{eqnarray}
<\rho> = \frac{3\,M_{\rm ap}}{4 \pi r_{t*}^3}.
\end{eqnarray}
These estimates are plotted in Figure~1. This is a conservative
estimate of the central density, as no assumption is made for the
detailed density profile shape, as it cannot be constrained on these
scaled from lensing observations. Note that the compact core radius
assumed above for galaxy scale subhaloes, taken to be 0.1 kpc, is not
used in the above estimate for the central density.

Considering X-ray clusters in the context of MOND, a massive neutrino
with a mass of $\sim$ 2 eV has been invoked by several authors to
explain observational data (Sanders 2003; Skordis et al. 2006).
Below, we derive an independent constraint from combined strong and
weak lensing data of clusters on the mass of such neutrinos using
estimates of phase space densities. Neutrino oscillation experiments
provide limits on the mass differences between the 3 species
($\nu_{e}, \nu_{\tau}, \nu_{\mu}$) of $\Delta m_{\nu} \sim
\sqrt{10^{-3}\,{\rm eV}^{2}}$ (Fukuda et al. 1998). Considering
massive neutrinos with masses well above $\Delta m_{\nu}$, the maximum
density of the neutrino halo can be estimated using phase space
arguments (Tremaine \& Gunn 1979). Comparison with the maximum phase
space density of a neutrino halo are evaluated via:
\begin{eqnarray}
{\rho_{\nu, max} \over {\rm  2.3 \times 10^{-5}\,M_{\odot}\,pc^{-3}}} 
\,(\frac{\sigma}{\rm 400\,km\,s^{-1}})^{-3} = M_{\nu}^4; \nonumber \\ 
M_{\nu}^4 =({m_{\nu_e} \over 2 eV})^4 + ({m_{\nu_{\mu}} \over 2 eV})^4 
+ ({m_{\nu_{\tau}} \over 2 eV})^4 + (\frac{m_1}{2 eV})^4 + (\frac{m_2}{2 eV})^4.
\end{eqnarray}
In computing the phase space density, we have used the formula of
Sanders (2003) to be conservative instead of Sanders (2007); the
latter gives a factor of three lower densities for the same neutrinos.
Here we have allowed for two species of sterile neutrinos. This has
been invoked to explain the results of neutrino oscillations detected
by LSND experiment with mass difference of $\sqrt{\Delta m^2} \sim
1$eV; e.g., two sterile neutrinos with masses $m_1=7$eV and
$m_2=6.9$eV.  Note the cosmic abundance of all neutrino species is
constrained by:
\begin{eqnarray}
{\Omega_{\nu_e}+ \Omega_{\nu_\mu}+ \Omega_{\nu_\tau} + \Omega_{\nu_s} \over 0.125 (H0/70)^{-2} }
= { m_{\nu_e}+ m_{\nu_\mu}+ m_{\nu_\tau} + m_{1} +m_{2} \over 6{\rm eV}}. 
\end{eqnarray}

The resultant phase space densities using Sanders (2003) are plotted as lines in
Figure~1. The dashed line corresponds to a Universe with just 3 species of
$m_{\nu}=2$eV ordinary (electron, muon and tau) neutrinos 
and a Universe with 0.08 eV ordinary and two sterile
neutrino species with masses of 7 eV and 6.9 eV respectively (solid line style).
\footnote{We do not
differentiate between the physical mass and the thermal mass since
they are nearly the same for eV range sterile neutrinos, which could
be produced non-thermally.}  The total non-baryonic neutrino fraction
$\Omega_{\nu_e}+ \Omega_{\nu_\mu}+ \Omega_{\nu_\tau} + \Omega_{\nu_s}$
ranges from 0.125 to 0.3 (the second model).  These numbers are in
broad agreement with the cosmology proposed by Skordis et al. (2006)
to account for CMB, without violating current limits on electron
neutrinos.  However, as seen in Figure~1, both possibilities fall
short of explaining the lensing data, in fact to explain the data
with neutrinos $\Omega_{\nu}\,>\,0.3$ is required.

\subsection{The baryonic contribution to cluster and cluster galaxy 
central densities}

The baryonic matter content of galaxy clusters is dominated by the
X-ray emitting intra-cluster gas. The gas mass exceeds the mass of
optically luminous material by a factor $\sim$ 6 (White et al. 1993;
Fukugita, Hogan \& Peebles 1998). As the emissivity of the X-ray
emitting gas is proportional to the square of its density, the gas
mass profile can be accurately determined from X-ray data. Measuring
the total mass profile is required to estimate the gas mass fraction,
and is more challenging to determine as it requires the direct
measurement of the gas temperature profile and the assumption of
hydrostatic equilibrium for the gas. Observations of nearby and
intermediate redshift clusters in the luminosity range ($L_{X,0.1-2.4}
\gtrsim 5 \times 10^{44} h_{70}^{-2}\,{\rm erg\,s^{-1}}$) and the
temperature range $kT\,>\,5\,{\rm keV}$, the average mass fraction in
stars (in galaxies and intra-cluster light combined) $f_{\rm star} \sim
0.16\,\sqrt{h_{70}}\,f_{\rm gas}$ (Lin \& Mohr 2004; Balogh et
al. 2001).

Fitting the results for a total of 68 clusters Allen et al. (2003;
2007) derive the gas fraction, $f_{\rm gas} = 0.1104 \pm 0.0016$.
These observational determinations of $f_{\rm gas}$ suggest that the
baryonic contribution to the central densities computed above are of
the order of $\sim 11\%$. This is also in agreement with recent
results from lensing and X-ray analysis reported in Takahashi \& Chiba
(2007). Therefore, our estimates of the central density from weak and
strong gravitational lensing reported above do primarily reflect the
dark matter density. The central density plotted in Figure~1 is now
corrected by the estimated factor of $0.89$ to reflect that of the
dark matter component alone.

Below we describe the estimation of the baryonic contribution to the
central density estimates for cluster galaxies. The results of the
maximum likelihood analysis in addition to providing a constraint on
the mass enclosed within an aperture also provide a constraint on the
total mass to light ratio. To separate the contribution of the
baryonic component, we estimate the stellar masses and subtract them
from the total aperture masses. To do so, the stellar mass-to-light
ratios are computed in the V-band. Using stellar population synthesis
models (Bruzual \& Charlot 2003), we estimate the stellar
mass-to-light ratios for template early-type galaxies at the redshifts
of these clusters. For the clusters studied here the ratio of the
total aperture mass-to-light ratio to the stellar mass-to-the light
ratio is the V-band is a factor of 2 - 3. Using these derived stellar
mass-to-light ratios and combining with the luminosity, we calculate
the total stellar mass within the aperture in these cluster
galaxies. We then subtract this from the total mass within the
aperture inferred from lensing. Doing so conservatively, we estimate
that on average at most $\sim 33\%$ of the contribution to the central
density derives from baryons. Using the computed stellar mass to light
ratios, we scale the values plotted in Figure~1 for cluster galaxies
in each cluster accordingly to derive the dark matter densities. We derive
the equivalent temperature (in keV) for clusters and cluster galaxies 
from their velocity dispersions via the relation:
\begin{eqnarray}
T\,=\left( {\sigma \over 400\,{\rm km\,s^{-1}}} \right)^2\,{\rm keV}.
\end{eqnarray}

In summary, we find that the central density of massive, lensing
clusters with well calibrated mass models precludes the possibility of
dark matter as $2\,{\rm eV}$ neutrinos that are required by
MOND. Moreover, the phase space density calculated for galaxies in
clusters also appears to be inconsistent with that of  two species of sterile 
neutrinos.  In general, our results are in line with $\Lambda$CDM
that postulates the existence of more massive dark matter particles
than neutrinos.

\section{Current limits on neutrino masses and ways out for MOND}

Have we detected the limiting phase space density of the dark matter
particle?  Is the particle inconsistent with neutrinos?  The mass of
ordinary neutrinos is still unknown although it must be non-zero. The
mass of electron neutrinos is measured in tritium $\beta$ decay
experiments. The decay results in a 3-helium, electron and an electron
anti-neutrino. If neutrinos have non-zero mass, the spectrum of the
electrons is deformed at the high energy part, i.e. the neutrino mass
determines the maximum energy of emitted electrons. To be exact, the
experiments measure the neutrino mass squared. Two running
experiments, Mainz and Troitsk, constrain the neutrino mass to be
above $m\,\sim\,0.05\,{\rm eV}$ and below $2.2\,{\rm eV}$. The upper
limit will get tighter once the KATRIN experiment starts in
2009-2010. The KATRIN experiment is expected to push the limit for
electron neutrino masses down by an order of magnitude. Our results
are barely consistent with neutrinos of such low mass. Therefore, the
cluster lensing data clearly rules out the possibility of these low
mass neutrinos constituting the bulk of dark matter as required by
MOND.

There are a few ways out for MOND to escape exclusion here:
\begin{itemize}

\item We have assumed negligible MOND corrections inside
clusters.  However, there could be regions where the $\mu$
is much smaller than unity, where classical MOND correction might be important.  
This is however a priori unlikely because clusters are in the strong or
moderate gravity regime with $g \ge a_0$ , hence $\mu = {g \over
g+a_0} \sim 0.5-1$.  

\item We have assumed a 1-to-1 history-independent relation between
mass distribution and gravity, as in GR and as in classical MOND.
This is not the case in recent co-variant incarnations of MOND, which
has (almost always) an additional fluid-like vector field in vacuum,
hence its stress energy tensor too bends the metric (Bekenstein 2004,
Sanders 2005, Zlosnik et al. 2007, Zhao 2007).  This fluid is
history-dependent, except in systems of equilibrium, like spiral
galaxies.  With this the ${\mathbf V}$ector-for-$\Lambda$ model (Zhao
2007) was able to match the $\Lambda$CDM cosmology, especially the
vacuum field can explain the tiny amplitude of the cosmological
constant $\Lambda \sim a_0^2/G$.  In these co-variant models, gravity
is determined by the instantaneous distribution of baryons and the
(dark) fluid; the latter tracks the former but with a phase-lag, hence
resembling the collisionless dark matter fluid.  We note that in the co-variant
V-Lambda fluid, lensing works as in GR and there are no anisotropic
stress corrections. This exit appears
plausible for MOND because classical MOND is non-covariant, and unless
it is given 3 or 4 vector degrees of freedom, lensing cannot be done
properly (i.e. getting the factor of two for light deflection, and
staying co-variant).

\item Neutrinos might have right-handed partners, sterile neutrinos,
whose mass is still poorly constrained by experiments, e.g., the
latest MiniBooNe experiment. It is foreseeable although not very
natural for sterile neutrinos with mass above $7\,{\rm eV}$ to be
partners to MOND gravity.

\item Aside from neutrinos it has been argued recently by Angus,
Famaey \& Buote (2007) that cluster dark matter could be baryonic in
the form of cold gas, analogous to the suggestion by Pfenniger \&
Combes (1994) on galaxy scales. However, there is no convincing
observational evidence to support this claim at the present time.

\end{itemize}

Nevertheless, what gives neutrino the tiny mass is still an unsolved
fundamental problem in physics, and in some theories it is linked to the
unsolved problem of dark energy (Mota et al. 2008).  Most recently it has been
proposed that MOND effects can come from a mass-varying neutrino with a
non-trivial coupling of the neutrino spin with the metric in Einsteinian
gravity (Zhao 2008 and references therein).   Neutrinos could cluster
significantly, there is tantalizing evidence for sterile neutrinos of 11eV in
the WMAP5 data (Angus 2008).

Examining the detailed feasibility of these options for a safe exit
for MOND are beyond the scope of this paper. In conclusion, using
combined strong and weak lensing data in clusters we constrain the
phase space density of the dark matter. Our current results rule out
neutrinos on eV scales as dark matter from lensing constraints derived
from galaxies in clusters on $\sim\,20\,-\,50\,{\rm kpc}$ scales.

Our constraints on phase space densities in the mild-acceleration
regime in galaxy cluster environments are much less-dependent on the
assumption of gravity in comparison to the case of isolated galaxies,
a regime where there could be appreciable MOND effects (Sellwood
2000).  Assuming GR, current data unfortunately does not yet suggest a
coherent picture of the true phase density of non-baryonic particles
that constitute dark matter. While the required finite core density of
dwarf spheroidal Galactic satellites favors sub-keV particles, such as
sterile neutrino-like warm dark matter (Gilmore et al. 2007), data
from the anomalous flux ratios of gravitational lensed radio quasars
(Miranda \& Maccio 2007; Metcalf \& Zhao 2002) and the flux
power-spectrum of SDSS Lyman-$\alpha$ systems favor cold dark matter
clumps, which would be erased by the streaming motions of sub-keV
sterile neutrinos or warm dark matter in general.  Our results suggest
that, in both MOND and in GR ordinary neutrinos and any sterile
neutrinos of $2 - 7\,{\rm eV}$ are insufficient to explain the
gravitational perturbations on scales of $\sim 20-50$ kpc that we
observe with galaxy cluster lensing data.

\section*{Acknowledgments} 

We thank the Dark Cosmology Centre, Niels Bohr Institute, for their
hospitality.  HSZ acknowledges partial support from UK STFC Advanced 
Fellowship and National Natural Science Foundation of China (NSFC under grant
No. 10428308. We acknowledge especially Benoit Famaey, Huanyuan Shan,
Martin Feix, Subir Sarkar and Marceau Limousin for helpful discussions
on MOND, neutrinos and lensing. Jerry Sellwood, Garry Angus and Benoit
Famaey are thanked for comments on the manuscript.


\begin{thebibliography}{}
\bibitem[]{} Allen, S, W., Rapetti, D., Schmidt, R., Ebeling, H.,
 Morris, G. \& Fabian, A. C., 2007, preprint, astro-ph/07060033
\bibitem[]{} Allen, S. W., Schmidt, R., Ebeling, H.,
 Morris, G. \& Fabian, A. C. \& van Speybroeck, L., 2004, MNRAS, 353,
457
\bibitem[]{}Angus G., 2008, preprint, arXiv:0805.4014
\bibitem[]{} Angus, G. W., Shan, H. Y., Zhao, H. S. \& Famaey, B., 2007a, ApJ, 654, L13
\bibitem[]{} Angus, G. W., Famaey, B., Buote, D. A., 2007b, preprint,
arXiv0709.0108 
\bibitem[]{} Balogh, M., Pearce, F., Bower, R., Richard, G. \& Kay, S., 2001,
MNRAS, 326, 1228
\bibitem[]{} Bradac, M., et al., 2006, ApJ, 652, 937
\bibitem[]{} Brainerd, T., Blandford, R. D., \&  Smail, I. R., 
1996, ApJ, 466, 623
\bibitem[]{} Broadhurst, T., Huang, X., Frye, B., \&  Ellis, R. S., 
2000, ApJ, 534, L15
\bibitem[]{} Broadhurst, T., Takada, M., Umetsu, K., Kong, X., 
Arimoto, N., Chiba, M., \&  Futamase,T., 2005, ApJ, 619, L143 
\bibitem[]{} Bruzual, G. \& Charlot, S., 2003, MNRAS, 344, 1000
\bibitem[]{} Clowe, D., Bradac, M., Gonzalez, A., Markevitch, M.,
Randall, S.,  Jones, C. \& Zaritsky, D., 2006, ApJ, 648, L109
\bibitem[]{} Courteau, S., Dutton, A., van den Bosch, F., MacArthur, L., Dekel, A., McIntosh, D \& 
Dale, D., 2007, ApJ, 671, 203
\bibitem[]{} Famaey, B., Gianfranco, G., Bruneton, J-P. \& Zhao, H., 2007,
Phys. Rev D., 75, 3002
\bibitem[]{} Famaey, B., Bruneton, J-P. \& Zhao, H. S., 2007, MNRAS, 377, L79
\bibitem[]{} Fukugita, M., Hogan, C. \& Peebles, J., 1998, ApJ, 503, 518
\bibitem[]{} Fritz, A., Ziegler, B., Bower, R., Smail, I \& Davies, R., 2005, MNRAS, 358, 233
\bibitem[]{} Geiger, B., \& Schneider, ., 1998, MNRAS, 295, 497
\bibitem[]{} Goerdt, T., Moore, B., Read, J., Stadel, J. \& Zemp, M., 2006,
MNRAS, 368, 1073
\bibitem[]{} Halkola, A., \& Seitz, S., 2007, ApJ, 656, 739
\bibitem[]{} Holden, B. P., et al., 2005, ApJ, 620, L83
\bibitem[]{} Jorgensen, I., Chiboucas, K., Flint, K., Bergmann, M., Barr, J \& Davies, R., 2006, ApJ, 639, L9
\bibitem[]{} Kleyna, J., Wilkinson, M., Evans, N. W. \& Gilmore, G., 2005,
ApJ, 630, L141 
\bibitem[]{} Klypin, A. \& Prada, F., 2007, preprint, arXiv0706.3554
\bibitem[]{} Kolb, E. \& Turner, M., 1990, The Early Universe, Addison-Wesley
Publishers, New York  
\bibitem[]{} Kneib, J.-P., Hudelot, P., Ellis, R. S., Treu, T., 
Smith, G.P., Marshall, P., Czoske, P., Smail, I. R., \& Natarajan, P., 
2003, Apj, 598, 804
\bibitem[]{} McGaugh, S. M., 2005, Phys. Rev Lett., 95, 171302
  N. \& Kim, J. H., 2007, ApJ, 659, 149
\bibitem[]{} Metcalf, B. \& Zhao, H. S., 2002, ApJ, 567, L5
\bibitem[]{} Miranda, M. \& Maccio, A., 2007, preprint, arXiv0706.0896 
\bibitem[]{} Limousin, M., Kneib, J-P., \& Natarajan, P., 2005, MNRAS, 356, 309
\bibitem[]{} Limousin, M., et al., 2007c, in press, ApJ, astro-ph/0612165
\bibitem[]{} Limousin, M., Kneib, J-P., Bardeau, S., Natarajan, P., 
Czoske, O., Smail, I., Ebeling, H., \& Smith, G. P., 2007a, A\&A, 461, 881
\bibitem[]{} Lin, Y-T \& Mohr, J., 2004, ApJ, 617, 879 
\bibitem[]{} Mota D.F., Pettorino V., Robbers G., Wetterich C., 2008, Phys.Lett. B, 663, 160
\bibitem[]{} Natarajan,P., \&  Kneib, J.-P., 1997, MNRAS, 287, 833 
\bibitem[]{} Natarajan, P., Kneib, J.-P., Smail, I. R., \& Ellis, 
R. S., 1998, ApJ, 499, 600 
\bibitem[]{} Natarajan, P., Kneib, J.-P., \&  Smail, I. R., 2002, ApJ, 580, L10
\bibitem[]{} Natarajan, P., \& Springel, V., 2005, ApJ, 617, L13
\bibitem[]{} Natarajan, P., De Lucia, G., \& Springel, V., 2007, MNRAS, 376, 180\bibitem[]{} Navarro, J. F., Frenk, C. S., \& White, S. D. M., 1997, ApJ, 490, 493
\bibitem[]{} Pahre, M., Djorgovski, G \&  de Carvalho, R., 1998, AJ, 116, 1591
\bibitem[]{} Pointecouteau, E. \& Silk, J., 2005, MNRAS, 364, 654
\bibitem[]{} Sanders, R. H. \& McGaugh, S. M., 2002, ARA\&A, 40, 263
\bibitem[]{} Sanders, R. H., 2003, MNRAS, 342, 901
\bibitem[]{} Sanders, R.H., 2005, MNRAS
\bibitem[]{} Schneider, P., Ehlers, J., \&  Falco, E., 1992, 
{\it Gravitational Lenses}, Springer-Verlag, Berlin.
\bibitem[]{} Schneider, P., \&  Rix, H.-W., 1997, ApJ, 474, 25 
\bibitem[]{} Sellwood, J. 2000, ApJ, 540, L1 
\bibitem[]{} Skordis, C., Mota, D. F., Ferreira, P. G. \& Boehm, C.,
  2006, Phys. Rev. Lett., 96, 1301
\bibitem[]{} Takahashi, R. \& Chiba, T., 2007, astro-ph/0701365, preprint
\bibitem[]{} Tremaine, S. \& Gunn, J., 1979, Phys. Rev. Lett., 42, 407
\bibitem[]{} Viel, M., Becker, G., Bolton, J., Haehnelt, M., Rauch, M. \& Sargent, W., 2007,
preprint, arXiv0709.0131
\bibitem[]{} Viel, M., Lesgourgues, J., Haehnelt, M., Matarrese, S. \& Riotto, A., 2006,
Phys. Rev. Lett., 97, 071301
\bibitem[]{} White, S. D. M., Navarro, J., Evrard, A. \& Frenk, C. S., 1993, Nature, 366, 429
\bibitem[]{} Wu, B. X., Zhao, H. S., Famaey, B., Gentile, G., Tiret, O., Combes, F., Angus, G. W. \& Robin, A. C., 2007, ApJ, 665, L101
\bibitem[]{} Xu, B. X., Wu, X. B. \& Zhao, H. S., 2007, ApJ, 664, 198
\bibitem[]{} Zhao, H. S., 2005, A\&A, 444, L25
\bibitem[]{} Zhao, H. S. \& Famaey, B., 2006, ApJ, 638, L9
\bibitem[]{} Zhao, H. S., 2007, ApJ, 671, 1 arXiv0710.3616
\bibitem[]{} Zhao, H.S. 2008, preprint, arXiv:0805.4046

\end{thebibliography}
\end{document}